
%
%
\input phyzzx
%
%
%
\def\sut{$ su_q(2) $ }
\def\suo{$ su_q(1,1) $ }
\def\sun{$ su_q(n) $ }
\def\ra{\rightarrow}
\def\cl{$ q \ra 1 $}
\def\hf{\textstyle{1 \over 2}}
\def\rep{representation }
\def\reps{representations }
\def\irrep{irreducible representation }
\def\irreps{irreducible representations }

\def\qiq{$ q \leftrightarrow q^{-1} $}
\def\RCNP{\address{Research Center for Nuclear Physics\break
        Osaka University, Ibaraki, Osaka 567, Japan}}
%
%
\REF\qg{There are many overview articles. Here we shall cite only three.
    \nextline
    M. Jimbo, in `Quantum Group and Quantum Integrable Systems', ed. by
    M.-L. Ge, World Scientific (1992) p1,
    \nextline
    P. P. Kulish, {\it ibid} p99,
    \nextline
    L. C. Biedenharn, in Lec. Notes Phys. {\bf 382} (1991) ed. by
    V. V. Dodonov and V. I. Man'ko, Springer-Verlag p147.}
\REF\ros{M. Rosso, Comm. Math. Phys.{\bf 117} (1988) 581.}
\REF\ru{V. Pasquier and H. Saleur, Nucl. Phys. {\bf B330} (1990) 523,
    \nextline
    G. Lusztig, Contemp. Math. {\bf 82} (1989) 237.}
\REF\kas{M. Kashiwara, Comm. Math. Phys. {\bf 133} (1990) 249.}
\REF\ku{P. P. Kulish, Theo. Matem. Fiz. {\bf 86} (1991) 157 (in Russian),
    \nextline
    Theor. Math. Phys. {\bf 86} (1991) 108 (English translation).
    \nextline
    See also, P. P. Kulish, ref. 1).}
\REF\zhe{A. S. Zhedanov, J. Phys. {\bf A25} (1992) L713.}
\REF\ai{N. Aizawa, `$ q \leftrightarrow q^{-1} $ Invariance of q-oscillator
        and new realizations of quantum algebras', to be published in
        J. Phys {\bf A}.}
\REF\qosc{L. C. Biedenharn, J. Phys. {\bf A22} (1989) L873, \nextline
    A. J. Macfarlane, J. Phys. {\bf A22} (1989) 4581, \nextline
    C.-P. Sun and H.-C. Fu, J. Phys. {\bf A22} (1989) L983, \nextline
    T. Hayashi, Comm. Math. Phys. {\bf 127} (1990) 129.}
%

\Pubtype={}
\Pubnum={RCNP preprint 052 }
\date={February 12, 1993}

\titlepage
\title{A Classically Singular Representation of $ su_q(n) $}
\author{N. Aizawa}
\RCNP

\abstract{A \rep of \sun, which diverges in the limit of \cl, is
investigated. This is an infinite dimensional and a non-unitary
\rep, defined for the real value of $ q, \ 0 < q < 1. $ Each \irrep
is specified by $ n $ continuous variables and one discrete variable.
This \rep gives a new solution of the Yang-Baxter equation, when the
R-matrix is evaluated.
It is shown that a continuous variables can be regarded as
a spectral parameter.}

\endpage

  When a physical realization of the quantum universal enveloping algebra
(quantum algebra)\refmark{\qg} is considered, its \rep theory plays a crucial
role. The \reps of the quantum algebra $ U_q({\bf \tt g}) $, where
$ {\bf \tt g} $ denotes a simple Lie algebra, are classified into three
categories according to the value of $ q $ : 1) generic, that is, $ q $ can
take any value except $ q = 0, \pm 1 $ and a root of unity, 2) $ q $ is a
root of unity, 3) $ q = 0 $.

  Rosso proved that for $ q $ generic, all finite dimensional \reps of
$ U_q({\bf \tt g}) $ are completely reducible and the irreducible ones are
classified in terms of highest weights.\refmark{\ros} In particular, they
can be regarded as deformation of the \reps of the classical
$ U({\bf \tt g}) $. When $ q $ is a root of unity, the \reps of
$ U_q({\bf \tt g}) $ become strikingly different from the classical case
\cl .\refmark{\ru} They are not completely reducible. Some finite dimensional
\reps are not the highest weight ones. The $ q = 0 $ case is called the
crystal base.\refmark{\kas}

  An infinite dimensional \rep of $ U_q({\bf \tt g}) $ is still an open
problem.

  In this paper, the \rep of \sun, which diverges in the classical limit,
is investigated. We call it a classically singular representation (CSR).
Such a \rep was first found by Kulish for the q-(bosonic) oscillator.
\refmark{\ku} Zhedanov found a q-oscillator realization of \sut and \suo
which diverges in the classical limit.\refmark{\zhe} Using the Fock \rep of
q-oscillator, he derived a infinite dimensional \rep of \sut and \suo. We
shall show that the another CSR of \sun (infinite dimensional) is possible.
This \rep gives a new solution of the Yang-Baxter equation, when the
R-matrix is evaluated.

  Recently, in connection with \qiq invariance of q-oscillator, a new
Jordan-Schwinger type realization of \sut was introduced.\refmark{\ai}
This realization reflects the fact that there exists a non-trivial central
element in the q-oscillator algebra, while the first proposed q-analogue of
the Jordan-Schwinger realization, ref.8), must assume the vanishing
central element. We generalize this realization to \sun, and applying
Kulish's CSR to it, we can obtain a CSR of \sun. Because the eigenvalue of
the central element is non-vanishing in Kulish's CSR of q-oscillator, the
Jordan-Schwinger realization of ref.8) does not work.

  The obtained CSR of \sun is infinite dimensional and non-unitary. Each
\irrep is specified by $ n $ continuous variables and one discrete one.
The continuous variables can take any positive real values and the discrete
one can take any integral (both positive and negative) numbers. These
\irreps can be regarded as follows : If the discrete variable is fixed,
we obtain a continuous family of \reps parametrized by the continuous
variables. In other words, a \irrep specified by the discrete variable
contains $ n $ parameters. By setting an appropriate condition, the
number of parameters reduces to one. We can regard this as a spectral
parameter, when the R-matrix is evaluated on the \rep space, although the
universal R-matrix of \sun contains no parameter except $ q $. We shall
discuss this point concretely for the case of \sut.

  We shall start with the definition of the q-oscillator algebra. It is
generated by three elements : q-creation operator $ a^{\dag} $,
q-annihilation operator $ a $ and number operator $ N $. They satisfy the
following relations,\refmark{\qosc}
$$
  \eqalign{
  & [N, a^{\dag}] = a^{\dag}, \qquad
    [N, a] = -a,
  \cr
  & aa^{\dag} - qa^{\dag}a = q^{-N}.}
                                                          \eqno(1)
$$
There exists a non-trivial central element,
$$
  C = q^{-N}([N] - a^{\dag}a),                            \eqno(2)
$$
where $ [N] \equiv (q^N - q^{-N})/(q - q^{-1}) $. This allow us to express
$ a^{\dag}a $ in terms of the operators $ N $ and $ C $,
$$
   a^{\dag}a = [N] - q^N C.                               \eqno(3)
$$
It should be noted that this is a direct consequence of the defining
relations (1), so more general than the standard prescription in which the
relation $ a^{\dag}a = [N] $ is assumed.

  Let us generalize the realization of \sut given in ref.7) to \sun.
Introducing $n$ independent (mutually commuting) q-oscillators,
$ a_i,\ a^{\dag}_i,\ N_i, \ i = $ 1, 2, $ \cdots, n$, we construct the
following $ 3(n-1) $ operators,
$$
  \eqalign{
  & X_i^+ = F_i^{-1/4} a^{\dag}_i\; F_{i+1}^{-1/4} a_{i+1}, \quad
    X_i^- = F_i^{-1/4} a_i\; F_{i+1}^{-1/4} a^{\dag}_{i+1},
  \cr
  & H_i = \hf (M_i - M_{i+1}),}
                                                          \eqno(4)
$$
where $ i = $ 1,2,$ \cdots, n-1, $ and the operator $ F_i $ and $ M_i $ are
defined by
$$
  \eqalign{
  & F_i \equiv 1 - (q-q^{-1})C,
  \cr
  & M_i \equiv N_i + { \ln \sqrt{F_i} \over \ln q}.}
                                                           \eqno(5)
$$
Using the relation (3), it is easy to verify that these operators form a
q-analogue of the Chevalley basis of \sun, that is, they satisfy the
following commutation relations.
$$
  [H_i, X_j^{\pm}] = \pm \hf A_{ij}\; X_j^{\pm}, \qquad
  [X_i^+, X_j^-] = \delta_{ij}[2H_i],
                                                            \eqno(6)
$$
where $ A_{ij} = 2\delta_{ij} - \delta_{i,j+1} - \delta_{i,j-1} $ is a
element of the Cartan matrix of $ su(n) $.

  Next, let us consider the CSR. In the following, we assume that $ q $ is
a real number. Kulish constructed two CSR's of the q-oscillator, one of
them is suitable for our purpose. It is given by,\refmark{\ku}
$$
  \eqalignno{
  & N |m> = m |m>, \qquad m = \cdots, -1, 0, 1, 2, \cdots, & (7.a)
  \cr
  & a^{\dag} |m> = L_{m+1}|m+1>,\ a|m> = L_m|m-1>,         & (7.b)
  \cr
  & C|m> = -q(v + \Delta)|m>,                             & (7.c)
  }
$$
where
$$
  L_m \equiv (q^{-m+1}v + q^{m+1}\Delta)^{1/2}, \quad
  v \equiv (1-q^2)^{-1}.                                   \eqno(8)
$$
The possible values of $ q $ and $ \Delta $ are restricted from
the requirement
that $ L_m^2 $ is non-negative,
$$
   0 < q < 1, \qquad \Delta > 0.                            \eqno(9)
$$
The peculiarity of this \rep is, 1) the matrix elements of $ a^{\dag},\ a $
and $ C $ diverge in the classical limit, since $ v \rightarrow \infty $
as \cl, 2) $ a^{\dag} $ is the hermite conjugate of $ a $, 3) there exist no
states which are annihilated by the q-annihilation or the q-creation
operator, 4) non-vanishing eigenvalue of the central element, 5) the CSR (7)
is irreducible for a fixed value of $ \Delta $, so we have an infinite
numbers of \irreps parametrized by $ \Delta $.

  In order to express the bases $ \{ |m> \} $ of (7) in terms of the
q-oscillator, we assume the existence of the state $ |0> $ such that,
$$
   N |0> = 0 ,\qquad C |0> = -q(v+\Delta) |0>,             \eqno(10.a)
$$
and for the definition of the norm
$$
   \eqalignno{
   & <0|0> = 1,                                              & (10.b)
   \cr
   & <0| a^m |0> = <0| (a^{\dag})^m |0> = 0,                 & (10.c)
   }
$$
where $ m $ is a positive integer. The bases $ \{ |m> \} $ can be written as,
$$
  \eqalignno{
  & |m> = ( \prod_{k=1}^m \; L_k )^{-1} (a^{\dag})^m |0>,   & (11.a)
  \cr
  & |m> = ( \prod_{k=-m+1}^0 \; L_k )^{-1} a^m |0>.         & (11.b)
  \cr
  & \hskip 7cm {\rm for}\ m \ \in {\bf Z_+}
  }
$$
These are orthonormal : $ <m|n> = \delta_{mn} $ for $ m,\ n, \in {\bf Z} $.
It is necessary to assume (10.c) to ensure the orthogonality of (11). These
bases disappear in the classical limit, since $ L_n $ diverges as \cl.

  It is now obvious that the CSR of \sun can be obtained from the
realization (4) and the CSR of the q-oscillator (7). First, note that the
fourth root of $ F_i $ appears in (4) and eigenvalue of $ F_i $ is negative :
$$
   F_i |m_i> = - {\Delta_i \over v} |m_i>.
$$
To avoid the multi-valuedness of the fourth root, we take the first branch.
It is easily verified that this choice do not lose generality. The \rep
space is the direct product of the ones of the CSR (7). Denoting its basis by
$$
  |\vec m> = |m_1,m_2,\cdots,m_n>
  = |m_1> \otimes |m_2> \otimes \cdots \otimes |m_n>,
$$
simple calculation gives the following \rep of \sun
$$
  \eqalignno{
  & X_i^+ |\vec m> = i \left( {v^2 \over \Delta_i \Delta_{i+1}} \right)^{1/4}
    L_{m_i + 1} L_{m_{i+1}}
    |m_1, \cdots, m_i+1, m_{i+1} -1, \cdots, m_n>
                                                              & (12.a)
  \cr
  & X_i^- |\vec m> = i \left( {v^2 \over \Delta_i \Delta_{i+1}} \right)^{1/4}
    L_{m_i} L_{m_{i+1}+1}
    |m_1, \cdots, m_i-1, m_{i+1} +1, \cdots, m_n>
                                                              & (12.b)
  \cr
  & H_i\; |\vec m> = \hf (m_i - m_{i+1} + { \ln(\Delta_i/\Delta_{i+1}) \over
         \ln q } )\; |\vec m>.                                  & (12.c)
  }
$$
In this \rep, $ X_i^+ $ and $ X_i^- $ are not hermite conjugate each other,
which stems from the choice of branch when the fourth root of $ F_i $ is
evaluated. It is possible, by taking the appropriate branch, to make
$ X_i^+ $ and $ X_i^- $ be hermite conjugate each other. In this case,
however, $ H_i $ become a non-hermitian matrix. In this sense, this \rep
is non-unitary.

  The \rep (12) is reducible. Because the action of $ X_i^{\pm} $ and
$ H_i $ preserves the sum $ \sum_{i=1}^n\;m_i $, the set
$ \{ |\vec m>\ |\ \sum_{i=1}^n\;m_i = m\} $ forms a invariant subspace for
each value of $ m = \cdots $, -1, 0, 1, 2, $ \cdots $. The action of
$ X_i^{\pm} $ and $ H_i $ also do not change the value of
$ \{ \Delta_i,\ i = 1, 2, \cdots, n \} $, so each \irrep is specified by
$ m $ and $ \Delta_i$'s. Since the values of $ m_i$'s are not bounded,
each \irrep is infinite dimensional. We can regard this situation as
follows : if the value of $ m $ is fixed, we have a continuous family of
\irreps parametrized by $ \Delta_i$'s.

  Let us investigate the case of \sut in more detail. The generators of
\sut are, $ J_+ = X_1^+,\ J_- = X_1^- $, and $ J_z = H_1 $. The central
element is given by,
$$
  C[su_q(2)] = [J_z][J_z -1] + J_+J_-.                       \eqno(13)
$$
Introducing the following notations,
$$
  \eqalignno{
  & J = \hf (m_1 + m_2), \qquad M = \hf (m_1 - m_2),         & (14.a)
  \cr
  & \kappa = (\Delta_1 \Delta_2)^{1/4}, \qquad \lambda = \Delta_1 / \Delta_2,
                                                             & (14.b)
  }
$$
and
$$
  |J,\; M,\; \kappa, \lambda> = |m_1> \otimes |m_2>,          \eqno(14.c)
$$
the CSR of \sut is rewritten as
$$
  \eqalign{
  & J_z |J,\; M,\; \kappa, \lambda> = (M + \hf {\ln \lambda \over \ln q})
        |J,\; M,\; \kappa, \lambda>,
  \cr
  & J_+ |J,\; M,\; \kappa, \lambda> = -i \sqrt v \;
        L_{J+M+1}^{(+)}\; L_{J-M}^{(-)} |J,\; M+1,\; \kappa, \lambda>,
  \cr
  & J_- |J,\; M,\; \kappa, \lambda> = -i \sqrt v \;
        L_{J+M}^{(+)}\; L_{J-M+1}^{(-)} |J,\; M-1,\; \kappa, \lambda>, }
                                                               \eqno(15)
$$
where
$$
  L_N^{(\pm)} = (q^{-N+1}v\kappa^{-1} + q^{N+1} \kappa \lambda^{\pm\hf})^{1/2}.
$$
The eigenvalue of the central element is specified by $ J $ and $ \kappa $ :
$$
  C[su_q(2)] |J,\; M,\; \kappa, \lambda> =
  - q^2 v ( q^{2J+1} \kappa^2 + q^{-2J-1}v^2 \kappa^{-2} + [2] v)
    |J,\; M,\; \kappa, \lambda>.                               \eqno(16)
$$
It is always negative. The possible values of $ J $ are (both positive and
negative) integers or half-integers. When $ J $ is an integer
(half-integer), $ M $ can take any (half) integral values. Each \irrep is
specified by $ J,\ \kappa $ and $ \lambda $.

  Here gives an interesting remark. A CSR of \suo is obtained from (15),
since the generators of \suo $ \{ K_{\pm},\ K_z \} $ can be expressed in
terms of \sut's,
$$
   K_{\pm} = i J_{\pm}, \qquad K_z = J_z.
$$
This procedure gives a unitary \rep of \suo. Another infinite dimensional
\rep , which connects smoothly to the classical one, is found in ref.7).

  Finally, the R-matrix is evaluated. Our convention of coproduct is,
$$
  \Delta(J_z) = J_z \otimes 1 + 1 \otimes J_z,  \quad
  \Delta(J_{\pm}) = J_{\pm} \otimes q^{-J_z} + q^{J_z} \otimes J_{\pm}.
                                                          \eqno(17)
$$
The universal R-matrix is given by,
$$
  R = q^{2J_z \otimes J_z} \sum_{l \ge 0} { (1-q^{-2})^l \over [l]! } \;
    q^{\hf l(l-1)}\; (q^{-J_z} J_-)^l \otimes (q^{J_z} J_+)^l.
                                                          \eqno(18)
$$
Before evaluating the matrix elements, we set $ \lambda = 1 $ (\ie
$ \Delta_1 = \Delta_2 $) in (15). Each \irrep is specified by $ J $ and
$ \kappa $. If $ J $ is fixed, a continuous family of \irrep, which is
parametrized by $ \kappa $, is obtained. The parameter $ \kappa $ can be
regarded as a spectral parameter, when the R-matrix is evaluated. Indeed, a
matrix element of (18) with respect to (15) reads,
$$
  \left(R(\kappa_1,\kappa_2)\right)_{M_1-l,M_2+l}^{M_1,M_2} =
  {1 \over [l]!} q^{\hf (2M_1-l) (2M_2+l) + l/2} \;
  \prod_{k=1}^l \Gamma_k^{(1)} \Gamma_k^{(2)},                \eqno(19)
$$
where
$$
  \eqalignno{
  & \Gamma_k^{(1)} = \{q^{-2J_1-1}v^2 \kappa_1^{-2} + q^{2J_1+1} \kappa_1^2 +
         v (q^{-2M_1-1+2k} + q^{2M_1 + 1 - 2k}) \}^{1/2},
                                                               & (20.a)
  \cr
  & \Gamma_k^{(2)} = \{q^{-2J_2-1}v^2 \kappa_2^{-2} + q^{2J_2+1} \kappa_2^2 +
         v (q^{-2M_2+1-2k} + q^{2M_2 - 1 + 2k}) \}^{1/2}.
                                                               & (20.a)
  }
$$
This R-matrix satisfies the Yang-Baxter equation with spectral parameters,
$$
 R_{12}(\kappa_1,\kappa_2) R_{13}(\kappa_1,\kappa_3) R_{23}(\kappa_2,\kappa_3)
 =
R_{23}(\kappa_2,\kappa_3) R_{13}(\kappa_1,\kappa_3) R_{12}(\kappa_1,\kappa_2).
                                                            \eqno(21)
$$
The R-matrix is also infinite dimensional and the difference property of
parameter :
$$ R(\kappa_1, \kappa_2) = R(\kappa_1 - \kappa_2) $$
does not hold in this case.

  The \rep discussed in this paper is peculiar to the quantum algebras,
since it has no classical counterpart. It will be easy to construct CSR for
other quantum algebras such as $ so_q(n),\ sp_q(n) $ etc. Another
possibility of \reps, which have no classical counterpart, is the one
vanishing in the limit of \cl. It will be interesting to consider such kind of
\rep.

  The author would like to thank Prof. P. P. Kulish for informing of his
work, ref.5). He also thanks Prof. M. Nomura and H. Sato for valuable
discussions.

\refout

\bye